\documentclass{article}
\usepackage[utf8]{inputenc}
\usepackage{graphicx}           
\usepackage{booktabs}
\usepackage{float}
\usepackage[colorlinks]{hyperref}

\usepackage[a4paper, total={6in, 8in}]{geometry}

\title{A Diamond Model Analysis on Twitter's Biggest Hack }
\author{Chaitanya Rahalkar}
\date{December 2021}

\begin{document}

\maketitle

\section{Abstract}
Cyberattacks have prominently increased over the past few years now, and have targeted actors from a wide variety of domains. Understanding the motivation, infrastructure, attack vectors, etc. behind such attacks is vital to proactively work against preventing such attacks in the future and also to analyze the economic and social impact of such attacks. In this paper, we leverage the diamond model to perform an intrusion analysis case study of the 2020 Twitter account hijacking Cyberattack. We follow this standardized incident response model to map the adversary, capability, infrastructure, and victim and perform a comprehensive analysis of the attack, and the impact posed by the attack from a Cybersecurity policy standpoint.

\section{Introduction}
\begin{figure}[h]
    \centering
    \includegraphics[width=\textwidth]{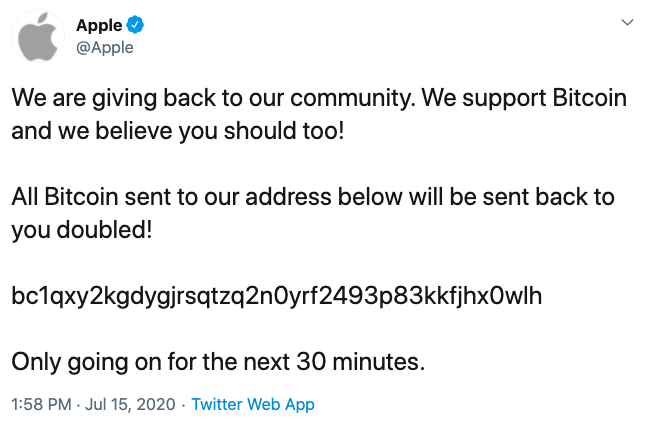}
    \caption{Apple's Official Twitter Account Promoting the Bitcoin Scam}
    \label{fig:tweet}
\end{figure}

On July 15, 2020,  between 20:00 UTC and 22:00 UTC, 130 Twitter profiles of highly influential agencies and people were compromised to promote a Bitcoin scam. High-profile entities like Elon Musk, Barack Obama, Bill Gates, Jeff Bezos, Kanye West, Kim Kardashian, Warren Buffet, etc. had their Twitter accounts compromised. This attack also included official corporate Twitter accounts of Apple, Uber, Cash App, etc. Of the total 130 compromised accounts, 45 were used to send out the scam Bitcoin Tweet \cite{twitter}\cite{isaca}. A screenshot of the Tweet is shown in \autoref{fig:tweet}. 
\\
\\
Vice, a Canadian-American magazine, was contacted anonymously by individuals claiming to be a part of the scam \cite{vice}. Vice reported that the attackers had managed to gain access to the internal administrative tools used by Twitter employees to alter the accounts and post the tweets on behalf of the users. 
\\
\\
The attack vector that was used to compromise the system was “social engineering". One or more (the actual number is unknown) Twitter employees were decisively tricked by the social engineering attack that took place on 15 July 2020. The attackers also managed to get past the 2FA (two-factor authentication) scheme that was added as an additional layer of protection to the employee accounts. As per the official report by Twitter, it was a phone spear-phishing attack. The report also stated that a successful attack required the attackers to gain access to the internal network as well as employee credentials that could grant them access to the internal support tools. Not all of the employees that were targeted had permissions to use the internal support tools. The attackers performed horizontal escalation to gain access to the internal systems through these obtained credentials and further penetrating into the internal infrastructure and finally taking over the Twitter accounts. Apart from just Tweeting from these 45 accounts, the attackers also accessed the DM (Direct Message) inbox of 36 accounts and downloaded the Twitter data for 7 of the accounts (using the Twitter data export feature).  
The Twitter support team uses these administrative tools to handle issues like account locking, Tweet content reviewing, and various other support-related problems. 
\\
\\
The primary motivation behind the attack was to extract money from users by leveraging highly influential people on Twitter. The scam tweet suggested that users should send Bitcoin to the Bitcoin address provided in the Tweet to get their sent money doubled, as a form of a charitable gesture. The attackers wisely used rich and influential individuals (and organizations) who could possibly engage in such a charitable action. Elon Musk, who was strongly encouraging people to invest in cryptocurrencies (especially Bitcoin) was also targeted in this attack which made it look legitimate. Close to \$110,000 had been deposited from 320 transactions in the Bitcoin wallet minutes after the Tweet was posted from the compromised accounts \cite{isaca}. Promptly, Twitter managed to remove the scam messages from these accounts. 

\section{Application of the Diamond Model}
\begin{figure}[h]
    \centering
    \includegraphics[width=0.75\textwidth]{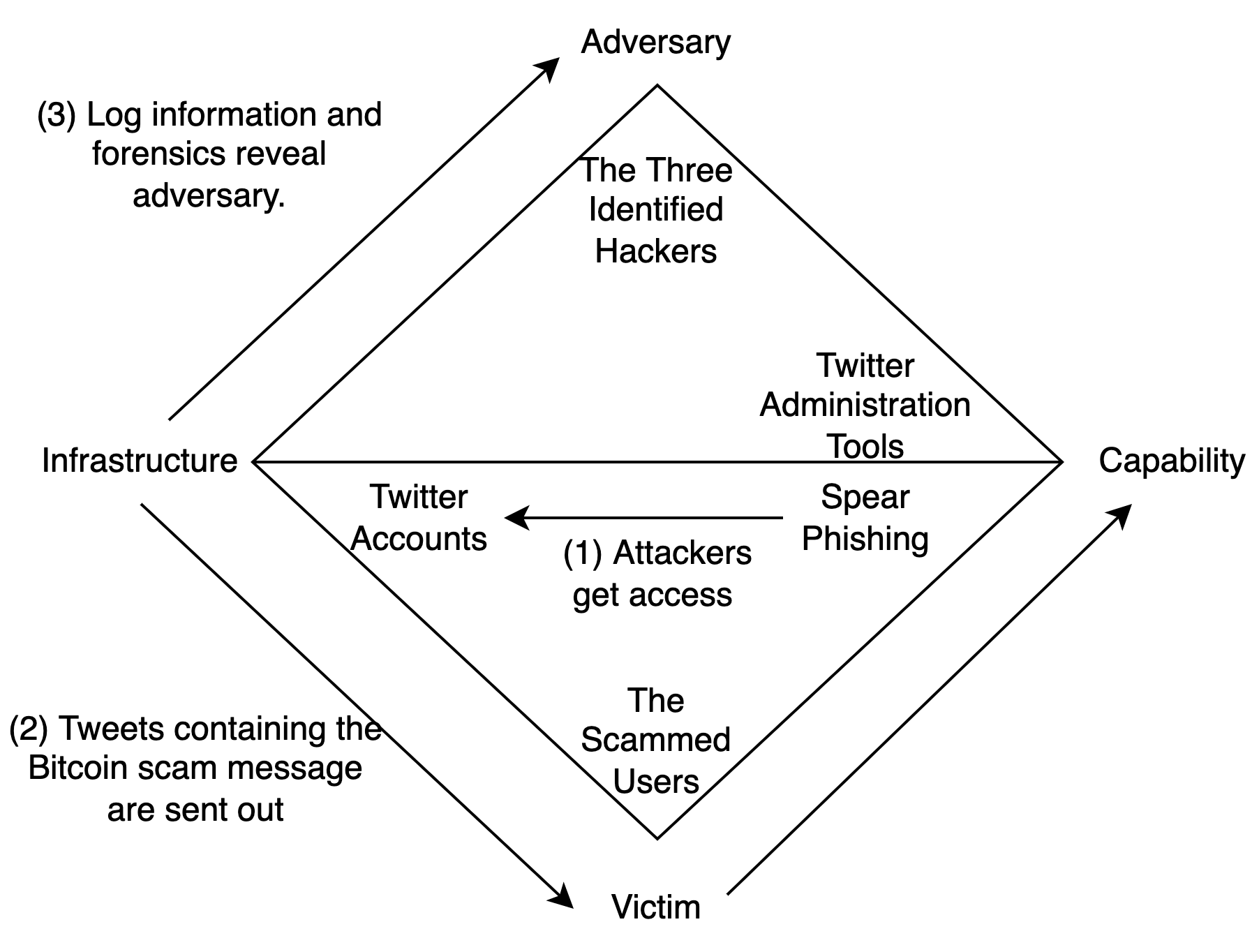}
    \caption{The Diamond Model Applied to the “Twitter Hack of 2020" Case Study}
    \label{fig:diamond}
\end{figure}

To better understand the fundamental aspects of the Cyberattack as well as develop a proactive scheme to discover, track and counter the activity and the adversary, we follow the Diamond model of intrusion approach. The Diamond Model does the job of integrating information assurance strategies and Cyber threat intelligence. It provides valuable insights by highlighting analytical opportunities and intelligence gaps. As per the paper, a Diamond Model's atomic element is an “event". For every event, we have four core features - “the adversary", “the capability" used by the adversary, “the infrastructure" leveraged using the capability, and “the victim" that was targeted to achieve the intended result. These core features are connected by edges. These edges signify a unique relationship between two features \cite{diamond}. \autoref{fig:diamond}
shows the diagrammatic representation of the Diamond model applied to the case study of the Twitter hack of 2020. 
\subsection{Adversary}
Post-mortem and further research analysis concluded that the scam was originated in the “OGUsers" group. “OGUsers" was a forum established for the trading of social media accounts. The FBI started an investigation on July 16 to look into the adversaries/perpetrators responsible for the scam. On July 31, the arrest of three individuals affiliated with the scam was announced by the United States Department of Justice \cite{justicegov}. These attackers had reportedly traded “original" (very old Twitter accounts with small, catchy usernames) for Bitcoin with the “OGUsers" group, to further expand the scam. Two individuals from Florida and one individual from the United Kingdom were charged and arrested. The three hackers with the intention of scamming people are mapped to the adversary feature in the diamond model. More specifically they are mapped to the “adversary operator" subcategory in the adversary feature. Here the “adversary customer" is the “adversary operator" since the hackers were the entities that were going to benefit from the Cyberattack. 
\subsection{The Capability}
Through a phone spear-phishing attack campaign targeted towards a specific set of Twitter employees, the attackers managed to gain access to a Twitter administrative tool (also known as the “Agent Tool") used by the Twitter support team. The attackers leveraged the capability of the administrative tool to arbitrarily set the email address used for resetting the password of any Twitter user. Security features like 2FA (two-factor authentication) were directly bypassed since the attackers were leveraging capabilities (password reset, changing user account emails, etc.) offered by the administrative tool. The attackers also managed to gain access to the Twitter team's Slack channel where information about remotely accessing internal network services and administrative tools was discovered. Since Twitter employees were working from home due to the ongoing COVID-19 pandemic, capabilities to remotely access internal network services were arranged for Twitter employees.  This insecure network design was used as an attack vector by the attackers to gain access to Twitter's internal systems. Therefore, “spear phishing" and “insecure network design" are mapped to “the capabilities" feature in the Diamond model. 
\subsection{The Infrastructure}
The attackers leveraged the Twitter administrative portal (also known as the “Agent Tool" in order to gain control over the users' Twitter accounts. As per the paper, the infrastructure feature is defined as the physical and/or logical communication structures the adversary uses to deliver a capability, maintain control of capabilities. The Twitter accounts were used to promote the Bitcoin scam campaign. The Twitter accounts were the logical communication structures used by the attackers to effect results (extract money through Bitcoin donations) from the victim users. The attackers used Twitter accounts of highly influential people to obfuscate the origin and attribution of the attack. From the perspective of the victims, these Twitter account owners would be considered adversaries. Therefore, the Twitter accounts would be distinctively categorized into a Type-2 Infrastructure. 
\subsection{The Victim}
The users who fell for the Bitcoin scam and sent money to the Bitcoin wallet expecting it to get doubled, are mapped to the “victim" feature of the model. More precisely, these users are sub-categorized into the “victim persona" feature. The Twitter users and organizations whose accounts were exploited to promote the scam were defamed in the process of the Cyberattack. Therefore, these people and organizations would also be indirect victims of the Cyberattack. 
\subsection{Meta Features}
The event meta-features are extended features of the Diamond model that represent the non-critical but important elements of Diamond events. 
Following are the meta-features considered for this case-study - 
\subsubsection{Timestamp}
The Cyberattack that compromised the Twitter accounts to send the Bitcoin scam message occurred on July 15, 2020, between 20:00 UTC and 22:00 UTC. 
\subsubsection{Phase}
The Cyberattack was conducted in a phased manner. This attack had a chain of events that were sequentially executed to achieve the end goal of the attackers. The attackers scraped data of Twitter employees from LinkedIn to study which employees had administrative privileges account-holder tools. Then they used the spear-phishing technique to get control over the employees' administrative accounts and then eventually compromised the users' accounts. 
\subsubsection{Methodology}
This meta-feature allows us to understand the general class of the activity. In this Cyberattack, spear-phishing emails were used to compromise the accounts of the Twitter employees.

\subsection{Activity Threads}
\begin{figure}[]
    \centering
    \includegraphics[width=\textwidth]{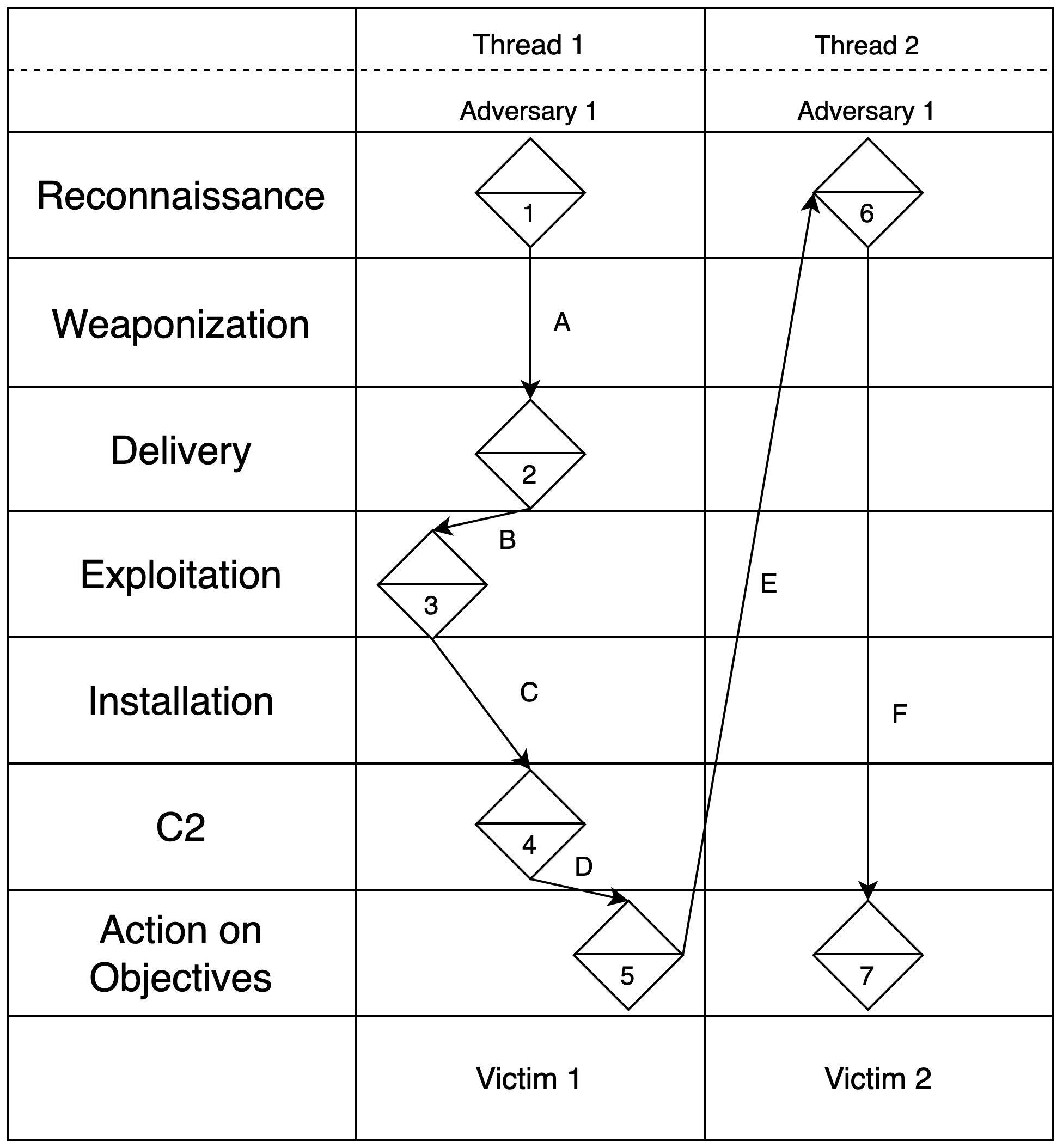}
    \caption{Activity Threads in the Diamond Model}
    \label{fig:activity}
\end{figure}

As per the paper, an activity thread is a directed phase-ordered graph where each vertex is an event and the arcs (i.e., directed edges) identify causal relationships between the events \cite{diamond}. We can map out two activity threads for the two victims identified in this case study. The users falling prey to this scam, and the defamed Twitter users whose accounts were used to propagate the scam are the two victims in this case. These two were the victims of the single adversary. One of them was a direct victim, the other was an indirect victim. We have two activity threads as shown in \autoref{fig:activity}. \autoref{tab:event} shows the event descriptions and \autoref{tab:arc} shows the arc descriptions for the case.

% \usepackage{booktabs}
\begin{table}[H]
\begin{tabular}{@{}lll@{}}
\toprule
Event & Hypothesis/Actual & Description \\ \midrule
1 & Actual & \begin{tabular}[c]{@{}l@{}}Attackers find out information about employees \\ from their LinkedIn accounts (to understand which of them had \\ administrative tools' access)\end{tabular} \\
2 & Actual & \begin{tabular}[c]{@{}l@{}}Attackers send a spear phished emails to these \\ employees and also bypass 2FA.\end{tabular} \\
3 & Actual & \begin{tabular}[c]{@{}l@{}}Attackers use these credentials to get access to the Slack \\ channel which has internal network access information.\end{tabular} \\
4 & Actual & \begin{tabular}[c]{@{}l@{}}Attackers manage to get into the internal network. \\ Using the administrative tools, they change the email \\ set for account password reset to an email address \\ which they control and gain access to Twitter accounts\\ of high profile users.\end{tabular} \\
5 & Actual & \begin{tabular}[c]{@{}l@{}}Attackers use these compromised Twitter accounts\\ to send out the Bitcoin scam Tweets.\end{tabular} \\
6 & Hypothesis & \begin{tabular}[c]{@{}l@{}}Attackers perform reconnaissance on which entities / \\ organizations' Twitter accounts to use to perform the\\ attack.\end{tabular} \\
7 & Actual & \begin{tabular}[c]{@{}l@{}}The attackers obscures their identity and use highly \\ influential entities / organizations to execute the attack, \\ and taint the reputation and trust of the entities / \\ organizations in the process.\end{tabular} \\ \bottomrule
\end{tabular}
\caption{Activity Thread Event Descriptions}
\label{tab:event}
\end{table}

% \usepackage{booktabs}
\begin{table}[H]
\begin{tabular}{@{}llll@{}}
\toprule
Arc & And / Or & Hypothesis / Actual & Provides \\ \midrule
A & And & Actual & Provides spear phishing targets \\
B & And & Actual & {[}None{]} \\
C & And & Actual & \begin{tabular}[c]{@{}l@{}}Provides internal network access\\ information\end{tabular} \\
D & And & Actual & \begin{tabular}[c]{@{}l@{}}Access to high profile Twitter \\ accounts\end{tabular} \\
E & And & Actual & {[}None{]} \\
F & And & Actual & {[}None{]} \\ \bottomrule
\end{tabular}
\caption{Activity Thread Arc Descriptions}
\label{tab:arc}
\end{table}
\section{Policy Assessment}
Cyberattacks have prominently increased in Cyberspace over the past decade. So far, the year 2020 has recorded the highest number of Cyberattack events that are publicly recorded \cite{breach}. The Twitter hack involved human as well as infrastructure-level negligence by the corporation. This is a very common attack vector used by Cyberattackers. Similar kinds of attacks could happen to other social media platforms like Facebook, Reddit, etc. to spread fraudulent / spam content. This case study is confined to the organizational layer in the Internet governance layers structure since it involves “Twitter" - a social media company and the users interacting with the platform.  As per the findings recorded in this paper, some Cybersecurity practices that Twitter (or generically any organization that wishes to protect their platform from such attacks) should incorporate in their existing policies are discussed in this section, to consider attacks conducted at such a scale and of such type.
\\
\\
Tackling these Cyberattacks at the “Organizational" layer can prevent further amplification of the attack that could affect the “National" and “Transnational" layers. The investigation report says that the attackers used the credentials of the employees (some of whom weren't authorized to access the administrative tools) and then vertically escalated to take control of the administrative tool. By performing correct network segmentation, the organization can prevent any kind of lateral movement. Therefore, even if the attackers manage to get hold of the credentials of the employees, if the employee does not have the right privileges, the attacker won't be able to move laterally and then potentially escalate to a higher privilege entity. Network segmentation isolates functionalities and users who are permitted to access these functionalities into distinct network segments. 
\\
\\
Incorporating zero-trust network policies is vitally important as well. In this case, the attackers impersonated the Twitter employees and managed to gain access to the internal network assets. However, by implementing a zero trust model, organizations won't have the traditional network edge that is found in an ordinary network infrastructure. It requires all users, whether inside or outside the organization's network perimeter to be authenticated, authorized, and validated for security configurations before granting access. Incorporating the Zero trust framework would not only make it very hard for Cyberattacker to perform lateral damage, but also it would protect them from such impersonation attacks that are simply based on credentials evaluation \cite{dfs}\cite{thompson}.

\section{Conclusion}
In this paper, we successfully applied the diamond model framework and studied the activity threads on the “Twitter Hack of 2020" case study, and analytically captured the events of the intrusion activity. We also documented the end-to-end process followed by the adversary to execute the Cyberattack and mapped the incidents to the correct “layers" of governance. We also proposed policy recommendations to make the intrusion analysis process for such kinds of Cyberattacks, streamlined from a Cybersecurity policy standpoint.

\end{document}